\documentclass{article}
\usepackage{spconf,amsmath,graphicx}
\usepackage{multirow}
\usepackage{booktabs}

\title{Fusing ASR Outputs in Joint Training for Speech Emotion Recognition}
\name{Yuanchao Li, Peter Bell, Catherine Lai}
\address{Centre for Speech Technology Research\\
        University of Edinburgh, Scotland, UK\\
        y.li-385@sms.ed.ac.uk, \{peter.bell, c.lai\}@ed.ac.uk}
\begin{document}
%
\maketitle
\begin{abstract}
Alongside acoustic information, linguistic features based on speech transcripts have been proven useful in Speech Emotion Recognition (SER). However, due to the scarcity of emotion labelled data and the difficulty of recognizing emotional speech, it is hard to obtain reliable linguistic features and models in this research area. In this paper, we propose to fuse Automatic Speech Recognition (ASR) outputs into the pipeline for joint training SER. The relationship between ASR and SER is understudied, and it is unclear what and how ASR features benefit SER. By examining various ASR outputs and fusion methods, our experiments show that in joint ASR-SER training, incorporating both ASR hidden and text output using a hierarchical co-attention fusion approach improves the SER performance the most. On the IEMOCAP corpus, our approach achieves 63.4\% weighted accuracy, which is close to the baseline results achieved by combining ground-truth transcripts. In addition, we also present novel word error rate analysis on IEMOCAP and layer-difference analysis of the Wav2vec 2.0 model to better understand the relationship between ASR and SER.
\end{abstract}
\section{Introduction}
\label{sec:intro}

Despite the rapid development of SER, its use in the wild remains difficult, due in part to the variability and vulnerability of human speech. To address this issue, researchers have proposed combining linguistic (i.e. transcript-derived) and acoustic information to improve the stability of SER. Traditional feature-level fusion and decision-level fusion, as well as the current tensor-level fusion trend, are widely used \cite{schuller2004speech}. Furthermore, many researchers have also compared the performance of various fusion methods \cite{sebastian2019fusion}. All of these works demonstrated the utility of combining linguistic information in SER.


Nonetheless, there is limited reliable text information in this area, particularly in real-life SER. First, when compared to ASR, the corpora sizes for SER are relatively small. For example, the popular IEMOCAP \cite{busso2008iemocap} only has about 12 hours of speech, whereas LibriSpeech \cite{panayotov2015librispeech} has about 1,000 hours. SER models built on such limited sized corpora don't generalize well to out-of-domain speech. Second, while previous studies proposed using ASR to generate transcripts for SER \cite{sahu19_interspeech}, ASR on emotional speech can often result in relatively high error rates.  Previous research has shown that emotion in speech degrades ASR performance, with emotional speech assumed to be a distortion of neutral speech  \cite{fernandez2004computational}. However, with the advancement of deep learning technologies, transfer learning for SER from ASR and joint training of ASR and SER have recently emerged \cite{zhou2020transfer, cai2021speech}. Nevertheless, the relationship between ASR and SER is still poorly studied, particularly what and how ASR features can benefit SER.

In this paper, we investigate various ASR outputs and fusion methods for a joint ASR-SER training model.
We analyze the Word Error Rate (WER) on IEMOCAP using Wav2vec 2.0 (W2V2), elucidating the reasons why ASR fails on emotional speech, which has not previously been reported. We also compare four ASR outputs from W2V2 and three fusion methods, and find that our proposed hierarchical co-attention fusion achieves 63.4\% Weighted Accuracy (WA), comparable performance to the baseline result using ground-truth transcripts.



\section{Related Work}
\label{sec:works}

The relationship between ASR and SER is an important but understudied topic. Although both tasks use speech signals as input, ASR works more at the frame level, whereas SER recognizes emotion on larger timescales.
Previous work \cite{fayek2016correlation} 
has demonstrated that features from the initial layers of both ASR and SER tasks are transferable and that the relevance of features gradually fades through deep layers. 
However, ASR features closer to the audio modality can outperform those closer to the text modality for arousal prediction and vice versa for valence prediction \cite{tits2018asr}. 

These findings suggest it is possible to use transfer learning for SER from ASR. 
For example, \cite{feng2020end} used hidden layer output from a pre-trained ASR model as linguistic features for SER and obtained promising results. Similarly, \cite{cai2021speech} used a single W2V2 model for both ASR and SER tasks and achieved State-Of-The-Art (SOTA) performance on 10-fold Cross-Validation (CV). However, gaps remain on foundational issues such as whether WER varies with emotion type, and where ASR features should enter SER models. In this paper, we build on this line of work, conducting new analyses on WER, ASR outputs, and fusion methods, with the goal of understanding how to best bring ASR into SER research.  

\section{Experiment Preparation and Preliminary Analysis}
\label{sec:analysis}

\subsection{Corpus and ASR Model}
In these experiments, we use the IEMOCAP corpus \cite{busso2008iemocap} as it is widely used in SER research. It consists of five dyadic sessions with ten actors, each with a scripted and improvised multimodal interaction. The corpus contains approximately 12 hours of speech that has been annotated by three annotators using ten emotion classes. We combine happy and excited, and use four categories: angry, happy, neutral, and sad, based on prior research \cite{li2019improved} . We removed utterances that did not have transcripts, bringing the total number of utterances used in this study to 5500.

We use the well-known W2V2 model \cite{baevski2020wav2vec} for the ASR model. W2V2 is a self-supervised learning framework composed of three major components: a CNN-based local encoder that extracts a sequence of embeddings from raw audio as latent representation, a Transformer network for context representation, and a quantization module for discretization. In this paper, we use the “Wav2vec2-base-960h” model, which has been fine-tuned using 960 hours of Librispeech.

\subsection{Preliminary Analysis of WER}
We calculate the WER on IEMOCAP to investigate the types of errors that occur in emotional speech (Table~\ref{wer}). Surprisingly, we see that the neutral WER is the second highest--we would expect WER for neutral speech to be the lowest as the least emotionally `distorted'. Furthermore, while Angry and Happy can have similarly variable prosodic characteristics (high f0 and intensity) \cite{paeschke2000prosodic} , Angry has the lowest WER while Happy has the highest. To determine the reasons for this, we conducted a word count analysis. The results in Table~\ref{wer} show that emotion classes with low WER typically have a low ratio of short utterances (word count less than 10).  
The fact that neutral speech is less prosodically variable may be the reason it has a lower WER than happy, even though its short utterances rate is the highest.

In fact, we calculated the WER based on the number of words, and found that the shorter the utterance, the higher the WER (see Table~\ref{count}). This finding contradicts a previous finding 
that longer utterances are more likely to have higher error rates
\cite{sahu19_interspeech}. This may be due to the fact that, when using W2V2, longer utterances contain more contextual information that the ASR can use to compensate for the negative effects of emotion. However, short utterances, and similarly disfluencies, are clearly very important in understanding affect in spoken dialogues \cite{tian2015emotion}. We hope that this preliminary analysis and novel findings will aid future ASR-SER research, by motivating the joint training to potentially learn these sorts of complex interactions.

\begin{table}
\centering
\caption{WER and word count analysis results. \emph{Utter} is the number of utterances of each emotion class, and \emph{Short\%} is the ratio of short utterances.}
\label{wer}
\begin{tabular}{lccccc}
\toprule
 & \textbf{Ang} & \textbf{Hap} & \textbf{Neu} & \textbf{Sad} & \textbf{Overall} \\ \midrule
\textbf{WER} & 22.8\% & 38.9\% & 36.3\% & 29.5\% & 32.7\% \\ \midrule
\textbf{Utter} & 1103 & 1615 & 1704 & 1078 & 5500 \\ \midrule
\textbf{Short\%} & 44.8\% & 55.7\% & 60.5\% & 52.7\% \\ \bottomrule
\end{tabular}
\end{table}

\begin{table}
\centering
\caption{WER according to different number of words N.}
\label{count}
\begin{tabular}{lccccc}
\toprule
\textbf{N} & \textbf{Ang} & \textbf{Hap} & \textbf{Neu} & \textbf{Sad} & \textbf{Overall} \\ \midrule
$\le$ \textbf{10} & 26.5\% & 54.8\% & 48.4\% & 47.2\% & 46.4\% \\ \midrule
\textbf{11}-\textbf{20} & 21.8\% & 37.5\% & 36.1\% & 30.4\% & 32.2\% \\ \midrule
\textbf{21}-\textbf{30} & 21.4\% & 31.0\% & 27.2\% & 22.6\% & 26.3\% \\ \midrule
$\ge$ \textbf{30} & 20.9\% & 31.0\% & 26.1\% & 20.1\% & 25.0\% \\ \bottomrule
\end{tabular}
\end{table}

\begin{figure*}[t]
  \label{model}
  \centering
  \includegraphics[width=0.97\textwidth]{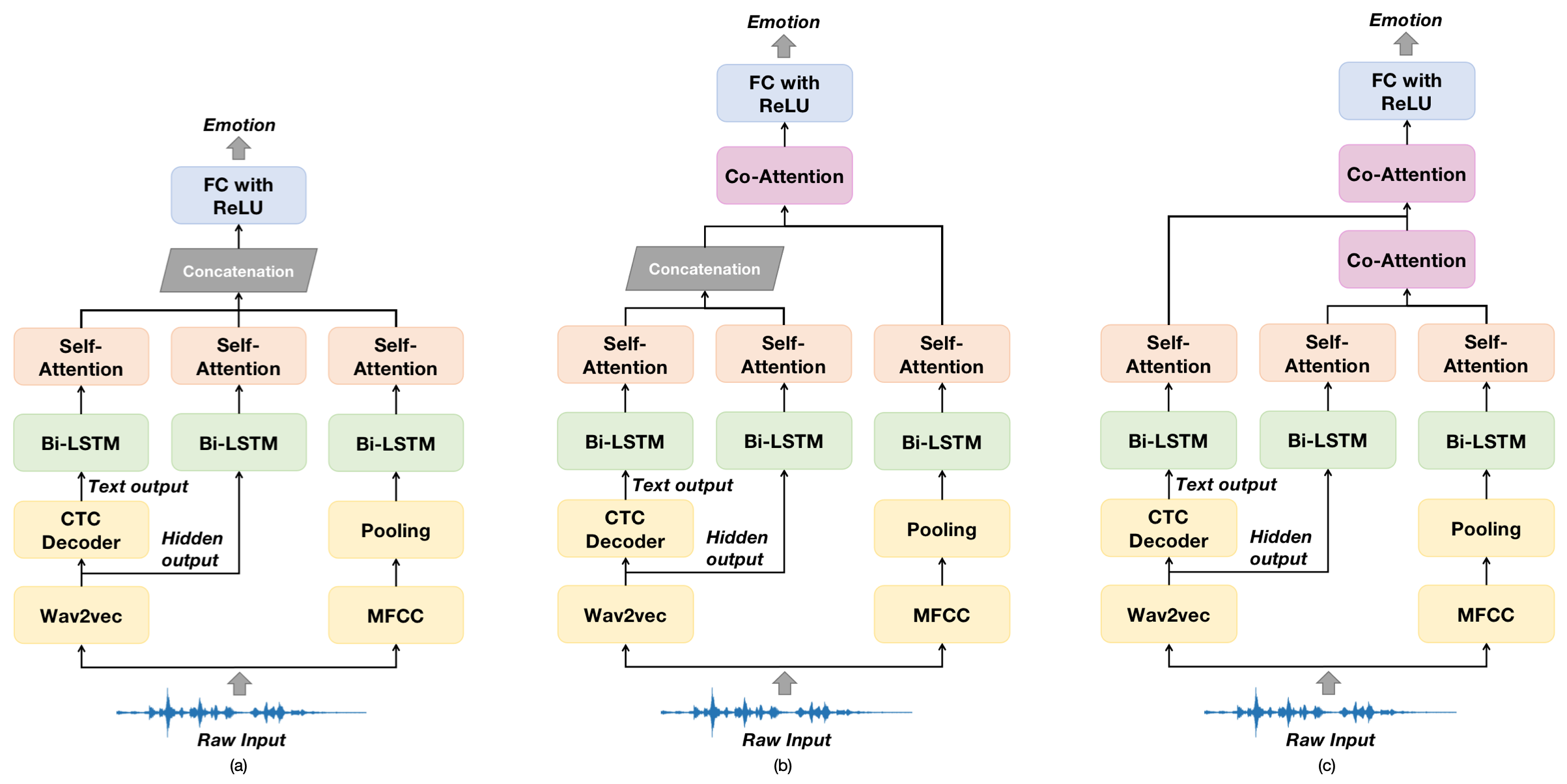}
  \caption{Proposed model with different fusion methods.}
  \label{fig:model-fusion}
\end{figure*}

\section{Method}
\label{sec:methods}

\subsection{Feature Extraction and Encodings}
In our model, the raw audio time-series is encoded by the W2V2 model (for ASR), and separately as MFCCs (for SER).    
On the ASR path, the W2V2 representations are then decoded to text by a word-level Connectionist Temporal Classification (CTC) decoder. Both the hidden state output and the text are extracted. A BERT model \cite{devlin2018bert} is then used to extract linguistic features from text output. On the SER path, a max pooling with kernel size 2 is conducted on the MFCC features to obtain an acoustic representation.

The ASR hidden output (short for hidden layer output), text output, and pooled MFCCs, are then encoded using 2-layer Bidirectional Long Short-Term Memory (Bi-LSTM) networks. Each layer contains 32 hidden units followed by a dropout with probability of 0.5. A self-attention layer with 16 heads and 64 nodes is applied on the output of the Bi-LSTM, which generates a fixed-length vector as the feature encoding. 

\subsection{Feature Fusion and Emotion Output}
We fuse the three encodings from the self-attention layer to produce a final vector for emotion classification. We compared three fusion methods (Fig.~\ref{fig:model-fusion}):  (a) concatenation, (b) concatenation with co-attention fusion, and (c) our proposed hierarchical co-attention fusion. The co-attention mechanism \cite{lu2019vilbert} concatenates two hidden-state vectors which exchange key-value pairs in self-attention, allowing features from one input channel to be incorporated into the other:

\begin{align}
    H_C &= Concat(H_{C_1}, H_{C_2}) \\
 H_{C_i} &= MultiHead(Q_{A}, K_{B}, V_{B})W^O \\
        &= Concat(head_1, ..., head_n) \\
 head_i &= Attention(Q_{A}W_i^Q,K_{B}W_i^K,V_{B}W_i^V)
\end{align}
where $W^O$, $W_i^Q$, $W_i^K$, and $W_i^V$ are trainable parameters. $Q_{A}$ represents the query from one input channel, while $K_B$ and $V_B$ represent the key and value from the other. The value of $n$ is 16, and $H_C$ denotes the final concatenated hidden states of co-attention.

Hierarchical approaches for emotion recognition have proved useful in fusing different-level or different-type of features in previous works \cite{tian2016recognizing}. Inspired by these works and the hierarchical characteristics of speech \cite{pascual2019learning}, we propose to fuse the input features from low to high level in a hierarchical manner using co-attention. Because MFCC features are extracted within frames, W2V2 features contain within-utterance context, and BERT features can learn cross-sentence context, we refer to them as frame-level, utterance-level, and dialog-level respectively, and hierarchically fuse them in this order to generate a fixed-length vector. This is passed to a fully connected output layer with \textit{Softmax} activation function to generate the probability distributions over emotion classes.

\begin{table*}
\centering
\caption{Performance comparison of using different models, features and fusion methods.}
\label{result}
\begin{tabular}{lllc}
\toprule
\textbf{Model} & \textbf{Feature} & \textbf{Fusion approach} & \textbf{WA} \\ \midrule
SER (\textit{baseline}) & Acoustic + Ground-truth transcripts & Concatenation & 63.5\% \\
 & & Co-attention & \textbf{64.9\%} \\ \midrule
ASR-SER & Acoustic + Hidden output (middle layer) & Concatenation & 58.7\% \\
 & & Co-attention & 59.1\% \\ \midrule
ASR-SER & Acoustic + Text output & Concatenation & 59.4\% \\
 & & Co-attention & 60.3\% \\ \midrule
ASR-SER & Acoustic + Hidden output (middle layer) + Text output & Concatenation & 62.2\% \\
 & & Co-attention & 62.9\% \\
 & & Hierarchical co-attention & \textbf{63.4\%} \\ \bottomrule
\end{tabular}
\end{table*}

\section{Experiments and Results}
\label{sec:experiments}
The model is optimized by the multi-task loss function:
\begin{align}
    \mathcal{L} &= \lambda \mathcal{L}_{\text{ASR}} + (1-\lambda) \mathcal{L}_{\text{SER}},
\end{align}
where $\mathcal{L}_{\text{ASR}}$ and $\mathcal{L}_{\text{SER}}$ are the losses for ASR and SER, respectively. We set $\lambda$ to 0.2, and use cross-entropy as the loss function. We use the Adam optimizer, with a learning rate of $10^{-4}$ and decay rate of $10^{-5}$. The gradients are clipped with a threshold of 5.0. The batch size is set to 20, and the number of epochs is limited to 100. We perform 5-fold CV and use WA to assess the performance. 

The experiment results are shown in Table~\ref{result}.
Here, we present (1) the ground-truth model incorporating human transcripts in the SER model (i.e. no ASR);  results of including (2) ASR hidden output or (3) text output; (4) the results of our proposed full model. We used the middle layer ASR output from W2V2 model because of its highest WA (see Table~\ref{layer}). 

As expected, the ground-truth model achieves the best performance as the ASR outputs contain recognition errors. Both ASR hidden output and text output help improve the SER performance over the acoustic features, although the difference is small. 
In general, the performance of co-attention fusion outperforms that of concatenation. It appears that the relatedness of two input channels is learned by attention. We also see that when using ground-truth transcripts, co-attention improves performance more than when using ASR outputs. It is plausible because the incorrectly recognized ASR features may have little or no relatedness to acoustic features, reducing the effectiveness of attention. Finally, our proposed approach: the joint ASR-SER model incorporating both ASR hidden output and text output using hierarchical co-attention achieves 63.4\% WA, which is nearly the same as the ground-truth transcript concatenation result and only 1.5\% less than the ground-truth co-attention result. This indicates that hierarchical fusion can help ameliorate ASR errors. 

\begin{table}
\centering
\caption{Layer-difference analysis results.}
\label{layer}
\begin{tabular}{llc}
\toprule
\textbf{Model} & \textbf{Feature} & \textbf{WA} \\ \midrule
SER (\textit{baseline}) & Acoustic & 51.4\% \\
 & Ground-truth transcripts & 60.0\% \\ \midrule
ASR-SER & Hidden output (first layer) & 45.6\% \\
 & Hidden output (middle layer) & 46.1\% \\
 & Hidden output (final layer) & 42.0\% \\
 & Text output & 47.7\% \\ \bottomrule
\end{tabular}
\end{table}


We conducted a layer-difference analysis study to determine which W2V2 layer contributes the most to the SER task. \cite{pasad2021layer} examined the layer-specific information in W2V2 intermediate speech representations and found that various acoustic and linguistic properties are encoded in different layers. As such, we compared the hidden-state outputs from the first layer (initial embeddings), middle layer, and final layer and found a surprising difference in SER performance. Table~\ref{layer} shows that the WA of using the hidden outputs from first and middle layers are close, but decreases significantly when using the final layer. Furthermore, we discovered that the text output outperforms the hidden outputs, which contradicts prior assertions that ``\textit{ASR features can be more robust than the text output of ASR}''
\cite{feng2020end}. This difference may be specific to W2V2 and requires further investigation, which we leave for future work.

We present previous works that also use ASR outputs in SER, but note that due to variation in experimental setup, results are not directly comparable. \cite{yoon2018multimodal} used the Google Cloud Speech API to generate transcripts that were then fused with MFCCs for SER. They achieved 69.1\% WA, likely due to their low WER (5.53\%). \cite{sahu19_interspeech} concatenated GloVe features obtained from Wit.ai API text transcription with acoustic features, obtaining a  62.9\% Unweighted Accuracy (UA) using an LSTM based model. \cite{feng2020end} jointly trained ASR-SER using log mel-scale filter bank outputs and decoder outputs from a pre-trained ASR model, and achieved 68.6\% WA. \cite{cai2021speech} used a single W2V2 as the training model for both ASR and SER rather than separating two tasks, and used 10-fold CV, which usually yields better results than 5-fold; \cite{zhou2020transfer} carefully fine-tuned the pre-trained ASR model on IEMOCAP before transferring the ASR features for SER. Note that since our goal is to analyze potential role of ASR in ASR-SER rather than achieve the SOTA performance, we did not focus on fine-tuning the model and parameters, but we expect that the performance of our model can be further improved by parameter tuning.

\section{Conclusions and Future work}
\label{sec:conclusions}
In this paper, we propose a joint ASR-SER training model that incorporates both ASR hidden output and text output using a hierarchical co-attention fusion approach. On IEMOCAP, it achieves 63.4\% WA, which is comparable to the ground-truth transcript model. By conducting WER analysis,
we demonstrate situations where ASR can fail on emotional speech. We hope that the findings, which have not previously been reported can contribute to understanding the ASR-SER relationship. In our future work, we will conduct detailed layer-wise analysis to understand what meanings the intermediate representations represent as the W2V2 layer becomes deeper. We will also try character-level or hybrid CTC decoders to solve the language model mismatch problem.

\section{Acknowledgements}
The authors would like to thank Dr. Richeng Duan from A$^{*}$STAR I2R and Han Feng from Huawei for valuable discussion.

\bibliographystyle{IEEEbib}
\bibliography{IEEE}

\begin{thebibliography}{10}

\bibitem{schuller2004speech}
Bj{\"o}rn Schuller, Gerhard Rigoll, and Manfred Lang,
\newblock ``{Speech emotion recognition combining acoustic features and
  linguistic information in a hybrid support vector machine-belief network
  architecture},''
\newblock in {\em {Proceedings of ICASSP 2004}}. IEEE, 2004.

\bibitem{sebastian2019fusion}
Jilt Sebastian, Piero Pierucci, et~al.,
\newblock ``{Fusion Techniques for Utterance-Level Emotion Recognition
  Combining Speech and Transcripts},''
\newblock in {\em Proceedings of Interspeech 2019}, 2019, pp. 51--55.

\bibitem{busso2008iemocap}
Carlos Busso, Murtaza Bulut, Chi-Chun Lee, Abe Kazemzadeh, Emily Mower, Samuel
  Kim, Jeannette~N Chang, Sungbok Lee, and Shrikanth~S Narayanan,
\newblock ``{IEMOCAP: Interactive emotional dyadic motion capture database},''
\newblock {\em {Language Resources and Evaluation}}, vol. 42, no. 4, pp.
  335--359, 2008.

\bibitem{panayotov2015librispeech}
Vassil Panayotov, Guoguo Chen, Daniel Povey, and Sanjeev Khudanpur,
\newblock ``{Librispeech: an asr corpus based on public domain audio books},''
\newblock in {\em {Proceedings of ICASSP 2015}}, 2015, pp. 5206--5210.

\bibitem{sahu19_interspeech}
Saurabh Sahu, Vikramjit Mitra, Nadee Seneviratne, and Carol Espy-Wilson,
\newblock ``{Multi-Modal Learning for Speech Emotion Recognition: An Analysis
  and Comparison of ASR Outputs with Ground Truth Transcription},''
\newblock in {\em {Proceedings of Interspeech 2019}}, 2019, pp. 3302--3306.

\bibitem{fernandez2004computational}
Raul Fernandez,
\newblock {\em {A computational model for the automatic recognition of affect
  in speech}},
\newblock Ph.D. thesis, Massachusetts Institute of Technology, 2004.

\bibitem{zhou2020transfer}
Sitong Zhou and Homayoon Beigi,
\newblock ``{A transfer learning method for speech emotion recognition from
  automatic speech recognition},''
\newblock {\em arXiv preprint arXiv:2008.02863}, 2020.

\bibitem{cai2021speech}
Xingyu Cai, Jiahong Yuan, Renjie Zheng, Liang Huang, and Kenneth Church,
\newblock ``Speech emotion recognition with multi-task learning,''
\newblock in {\em Proceedings of Interspeech 2021}, 2021.

\bibitem{fayek2016correlation}
Haytham~M Fayek, Margaret Lech, and Lawrence Cavedon,
\newblock ``{On the Correlation and Transferability of Features Between
  Automatic Speech Recognition and Speech Emotion Recognition},''
\newblock in {\em Proceedings of Interspeech 2016}, 2016, pp. 3618--3622.

\bibitem{tits2018asr}
No{\'e} Tits, Kevin El~Haddad, and Thierry Dutoit,
\newblock ``{ASR}-based features for emotion recognition: A transfer learning
  approach,''
\newblock in {\em Proceedings of Grand Challenge and Workshop on Human
  Multimodal Language (Challenge-{HML})}. 2018, pp. 48--52, Association for
  Computational Linguistics.

\bibitem{feng2020end}
Han Feng, Sei Ueno, and Tatsuya Kawahara,
\newblock ``{End-to-End Speech Emotion Recognition Combined with
  Acoustic-to-Word ASR Model},''
\newblock in {\em INTERSPEECH}, 2020, pp. 501--505.

\bibitem{li2019improved}
Yuanchao Li, Tianyu Zhao, and Tatsuya Kawahara,
\newblock ``{Improved End-to-End Speech Emotion Recognition Using Self
  Attention Mechanism and Multitask Learning},''
\newblock in {\em Proceedings of Interspeech 2019}, 2019, pp. 2803--2807.

\bibitem{baevski2020wav2vec}
Alexei Baevski, Yuhao Zhou, Abdelrahman Mohamed, and Michael Auli,
\newblock ``Wav2vec 2.0: A framework for self-supervised learning of speech
  representations,''
\newblock in {\em Advances in Neural Information Processing Systems}, 2020, pp.
  12449--12460.

\bibitem{paeschke2000prosodic}
Astrid Paeschke and Walter~F Sendlmeier,
\newblock ``{Prosodic characteristics of emotional speech: Measurements of
  fundamental frequency movements},''
\newblock in {\em ISCA Tutorial and Research Workshop on Speech and Emotion},
  2000.

\bibitem{tian2015emotion}
Leimin Tian, Johanna~D Moore, and Catherine Lai,
\newblock ``Emotion recognition in spontaneous and acted dialogues,''
\newblock in {\em Proceedings of ACII 2015}. IEEE, 2015, pp. 698--704.

\bibitem{devlin2018bert}
Jacob Devlin, Ming-Wei Chang, Kenton Lee, and Kristina Toutanova,
\newblock ``{BERT: Pre-training of deep bidirectional transformers for language
  understanding},''
\newblock {\em Proceedings of NAACL-HLT 2019}, pp. 4171--4186, 2019.

\bibitem{lu2019vilbert}
Jiasen Lu, Dhruv Batra, Devi Parikh, and Stefan Lee,
\newblock ``{ViLBERT: Pretraining task-agnostic visiolinguistic representations
  for vision-and-language tasks},''
\newblock in {\em Advances in neural information processing systems}, 2019.

\bibitem{tian2016recognizing}
Leimin Tian, Johanna Moore, and Catherine Lai,
\newblock ``Recognizing emotions in spoken dialogue with hierarchically fused
  acoustic and lexical features,''
\newblock in {\em Proceedings of SLT 2016}. IEEE, 2016, pp. 565--572.

\bibitem{pascual2019learning}
Santiago Pascual, Mirco Ravanelli, Joan Serra, Antonio Bonafonte, and Yoshua
  Bengio,
\newblock ``Learning problem-agnostic speech representations from multiple
  self-supervised tasks,''
\newblock in {\em Proceedings of Interspeech 2019}, 2019.

\bibitem{pasad2021layer}
Ankita Pasad, Ju-Chieh Chou, and Karen Livescu,
\newblock ``Layer-wise analysis of a self-supervised speech representation
  model,''
\newblock in {\em Proceedings of ASRU 2021}, 2021.

\bibitem{yoon2018multimodal}
Seunghyun Yoon, Seokhyun Byun, and Kyomin Jung,
\newblock ``Multimodal speech emotion recognition using audio and text,''
\newblock in {\em 2018 IEEE Spoken Language Technology Workshop (SLT)}. IEEE,
  2018, pp. 112--118.

\end{thebibliography}

\end{document}